\documentclass[prl,aps,showpacs,twocolumn,superscriptaddress,floatfix]{revtex4}

\usepackage{graphicx}
\usepackage{epsfig}
\usepackage{amsfonts}
\usepackage{amsmath,amssymb}
\usepackage{bm}

\begin{document}

\title{Isospin mixing in nuclei within the nuclear density functional theory}

\author{W. Satu{\l}a}
\affiliation{Institute of Theoretical Physics, University of Warsaw, ul. Ho\.za
69, 00-681 Warsaw, Poland}

\author{J. Dobaczewski}
\affiliation{Institute of Theoretical Physics, University of Warsaw, ul. Ho\.za
69, 00-681 Warsaw, Poland}
\affiliation{Department of Physics, P.O. Box 35 (YFL),
FI-40014 University of Jyv\"askyl\"a, Finland}

 \author{W. Nazarewicz}
\affiliation{Department of Physics \&
  Astronomy, University of Tennessee, Knoxville, Tennessee 37996}
\affiliation{Physics Division, Oak Ridge National Laboratory, P.O. Box
  2008, Oak Ridge, Tennessee 37831}
\affiliation{Institute of Theoretical Physics, University of Warsaw, ul. Ho\.za
69, 00-681 Warsaw, Poland}

 \author{M. Rafalski}
\affiliation{Institute of Theoretical Physics, University of Warsaw, ul. Ho\.za
69, 00-681 Warsaw, Poland}

\date{\today}

\begin{abstract}
We present  the self-consistent, non-perturbative  analysis of isospin
mixing using  the nuclear density functional approach and the
rediagonalization of the  Coulomb interaction in the good-isospin basis.
The largest isospin-breaking effects are predicted for $N$=$Z$ nuclei
and they quickly fall with the neutron excess. The unphysical isospin
violation on the mean-field level, caused by the neutron excess, is
eliminated by the proposed method.
We find a significant dependence of
the magnitude of isospin breaking  on the parametrization of
the nuclear interaction term. A  rough correlation has been found
between the isospin mixing parameter  and the difference of  proton
and neutron rms radii. The theoretical framework   described in this
study is well suited  to describe a variety of phenomena associated with
isospin violation  in nuclei, in particular the
isospin symmetry-breaking corrections to superallowed Fermi beta decays.
\end{abstract}

\pacs{21.10.Hw, % Spin, parity, and isobaric spin
21.60.Jz, %	Nuclear Density Functional Theory and extensions (includes HartreeÐFock and random-phase approximations)
21.30.Fe, % Forces in hadronic systems and effective interactions
23.40.Hc %	Beta decay, relation with nuclear matrix elements and nuclear structure
}
\maketitle

The isospin symmetry, introduced by Heisenberg \cite{[Hei32]} and Wigner
\cite{[Wig37]}, is largely preserved by strong interactions; a small
violation of isospin on the hadronic level is due to the difference in
the masses of the up and down quarks \cite{[Mil06]}. In atomic nuclei,
the main source of isospin breaking is the electromagnetic interaction
\cite{[Bli62],[Ber72]}.
Since the isovector and isotensor parts of
electromagnetic force are much weaker than the
strong interaction between nucleons, many effects associated with isospin
breaking in nuclei be can treated in a perturbative way. With this caveat,
the formalism of isotopic spin is  a very powerful concept in nuclear
structure and reactions \cite{[Wil69],[War06]}, where many spectacular
examples of isospin symmetry can be found.

The main effect of Coulomb force in nuclei is to exert a long-range
overall polarization effect on nuclear states whose detailed structure
is  dictated by the short-ranged strong force. The net effect of such a
polarization is a result of two competing trends: the nuclear force is
strongly attractive in the isoscalar neutron-proton channel, while the
Coulomb force acts against this attraction by making neutron and proton
states different. In order to explain this interplay, self-consistent
feedback  between strong and electromagnetic fields must be considered
to best locate the point of the nuclear equilibrium.

An
excellent example of this interplay is the systematic behavior of nuclear binding
energies: with increasing mass number, the  stability line bends away
from the $N$=$Z$ line towards the neutron-rich nuclei. The effect of
electromagnetic force on nuclear binding is clearly non-perturbative.
%without Coulomb term the binding energy of $^{208}$Pb of 1635\,MeV
%would have been about 800\,MeV larger.
Even in medium-mass nuclei, which
are of principal interest in this study, energy
balance between strong and Coulomb forces
is not tremendously favorable, e.g., 342\,MeV versus\ 72\,MeV in
$^{40}$Ca. The situation becomes dramatic in  superheavy
nuclei and in the neutron star crust, where  not only the binding but
also spectra are strongly impacted by the  Coulomb frustration effects
resulting from a self-consistent, non-perturbative  feedback between
strong and electromagnetic parts of the nuclear Hamiltonian
\cite{[Cwi96s],[Hor05s]}.

The strong motivator for studies of isospin breaking is nuclear beta decay. The new data in superallowed 0$^+$$\rightarrow$0$^+$ nuclear beta decays \cite{[Har05a]}, including  a number of
high-precision Penning-trap measurements, require improved calculations
of isospin-breaking corrections \cite{[Tow08],[Mil08]}. As far as  nuclear spectroscopy is concerned, there has been an increased interest in isospin-related phenomena in
recent years \cite{[War06]}. For instance, studies of excited states of
proton-rich nuclei with $N<Z$
resulted in significantly improved  information on Coulomb energy
differences \cite{[Ben07a]}. In some cases, observed Coulomb shifts
turned out to be surprisingly large \cite{[Ekm04]},
thus fueling speculations of  significant  nuclear charge-symmetry-breaking forces.
%In this context, recent self-consistent calculations have demonstrated
%that the many-body response against electrostatic polarization appears to be
%strongly sensitive to the choice of the nuclear effective interaction, in
%particular  the spin-dependent terms \cite{[Sto06s]}.

A precise description of Coulomb effects in nuclei constitutes a
notoriously difficult computational challenge.  In the shell-model
approach to the isospin mixing  (see,
e.g.,  Refs. \cite{[Orm85],[Orm95a]}), the effective shell-model
Hamiltonian including the Coulomb interaction  is diagonalized in a
proton-neutron basis to  account for non-perturbative effects. The
overall strength of the isospin-breaking interactions is usually
renormalized by  reproducing the rms proton point radii obtained from
spherical Hartree-Fock (HF) calculations \cite{[Orm95a]} or by fitting the
experimental isobaric  mass shifts \cite{[Orm85],[Ben99c]}. To take into account
the  coupling to the giant monopole  resonance that appreciably
influences the radial mismatch between the proton and neutron wave
functions \cite{[Ber72]}, single-particle wave functions can be taken from  HF
calculations. More precise treatments require determining the effective
Coulomb interaction  in the large space, which is possible in the
no-core shell model. Such calculations have been carried out for
$^{10}$C \cite{[Cau02x]} in the space allowing all 8\,$\hbar\Omega$
excitations relative to the unperturbed ground state. Currently, however,
 {\it ab-initio} approaches to superallowed Fermi transitions do not go beyond
$^{10}$C which marks the state of the art.

In heavier nuclei, especially those involving many nucleons outside
closed shells, the isospin mixing can be well described  by the
mean-field (MF) or energy-density-functional (EDF) methods
\cite{[Ben03s]}, where the Coulomb force  amounts to making the neutron
and proton single-particle orbitals  different, and the long-range
polarization effects (e.g., those related to the isoscalar and isovector
monopole resonance) are fully taken into account.

The fact that the MF
methods allow for precise treatment of  long-range operators is, in fact,
essential for the  physics of isospin mixing.
However, it was very early realized
\cite{[Eng70],[Bri70],[Cau80],[Cau82],[Aue83]} that these nice physical
properties of the MF methods are accompanied by unwanted spurious
effects related to the fact that the neutron and proton
single-particle states in $N\neq{}Z$ nuclei are different even without
Coulomb interaction included. Indeed, the presence of the neutron or
proton excess automatically yields isovector mean fields, i.e., different HF potentials for
protons and neutrons. This unwelcome feature has hampered
MF calculations of the isospin mixing beyond the $N=Z$ systems (see,
e.g., Ref.~\cite{[Dob95a]}). To overcome this difficulty in the present study, we employ the
mean-field methods in the framework of Refs.~\cite{[Cau80],[Cau82]},
which is entirely free of the spurious isospin mixing. Thereby, for
the first time, we determine the isospin mixing within the context of
modern EDF methods.

We begin by noting that the self-consistent MF state $|\textrm{MF}
\rangle$ can be expanded in good-isospin basis $|T,T_z\rangle$:
\begin{equation}\label{mix}
|\textrm{MF} \rangle = \sum_{T\geq |T_z|}b_{T,T_z}|T,T_z\rangle,
\quad \sum_{T\geq |T_z|} |b_{T,T_z}|^2 = 1,
\end{equation}
where $T$ and $T_z=(N-Z)/2$ are the total isospin and its third
component, respectively.
%, and $\sum_{T\geq |T_z|} |b_{T,T_z}|^2 = 1$.
The basic assumption behind our approach is that the states
$|T,T_z\rangle$
capture  the right
balance between strong and Coulomb interactions; i.e., they
contain self-consistent polarization effects to all orders. Below we
shall validate this assumption by varying the MF charge $e_{\text{MF}}$,
which defines the strength of the Coulomb interaction at the MF
level, that is, when solving the self-consistent HF  equations. On the
other hand, the mixing coefficients $b_{T,T_z}$ are not reliably
determined by the MF method, because they are affected by the
neutron-excess-induced spurious isospin mixing.

To assess the true isospin mixing,
the total Hamiltonian ${\hat H}$ (strong interaction plus the Coulomb interaction with the physical charge $e$) is rediagonalized  in the
space spanned by the good-isospin wave functions:
\begin{equation}\label{mix2}
|n,T_z\rangle = \sum_{T\geq |T_z|}a^n_{T,T_z}|T,T_z\rangle ,
\end{equation}
where $n$ enumerates the  eigenstates $|n,T_z\rangle$ of ${\hat H}$. The value of  $n=1$, corresponds to the isospin-mixed  ground
state (g.s.). In the following, the g.s.\ isospin-mixing parameter
$\alpha_C = 1- |a^{n=1}_{|T_z|,T_z}|^2$ and energy  $E_{n=1,T_z}$
obtained after rediagonalization (AR) are distinguished  from the quantities
$\alpha_C = 1- |b_{|T_z|,T_z}|^2$ and
$E_{|T_z|,T_z}=\langle T\mbox{=}|T_z|,T_z|\hat{H}|T\mbox{=}|T_z|,T_z\rangle$,
obtained before rediagonalization (BR; isospin projection after variation).
The AR results are superior to the BR results as they are based on the double variational principle.

Our self-consistent calculations have been  carried out by using the
SLy4  EDF parameterization
\cite{[Cha95s]} and  the HF solver  {\sc HFODD} \cite{[Dob04c]} that
allows for arbitrary spatial deformations of intrinsic states. Both direct and
exchange Coulomb terms  are calculated exactly. We adopted the standard
technique of isospin projection  \cite{[Rin80]}. Details pertaining to  our
method   can be found in Ref.~\cite{[Raf09]}, together with numerical tests.

\begin{figure}
\includegraphics[angle=0,width=0.7\columnwidth,clip]{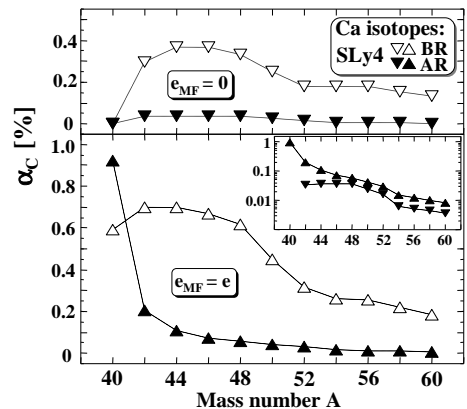}
\caption[T]{\label{fig1}
Isospin-mixing parameter $\alpha_C$ for the even-even  Ca isotopes determined before
(BR) and after (AR) rediagonalization in the good-isospin basis
$|T,T_z\rangle$. The basis states were generated by means of self-consistent calculations
without ($e_{\text{MF}}$=0; upper panel) and with ($e_{\text{MF}}$=$e$; lower panel) the Coulomb term. The inset shows the AR results
plotted in the logarithmic scale.
}
\end{figure}
To illustrate the effect of the spurious isospin mixing, in Fig.~\ref{fig1} we show the BR and AR results  for the even-even  Ca
isotopes.
Without Coulomb interaction ($e_{\text{MF}}=0$), there is no
isospin mixing in the $N$=$Z$ nucleus $^{40}$Ca, but the
neutron-excess-induced  mixing  appears in
all systems with
$N\neq{}Z$. The BR spurious mixing is
quite large, $\alpha_C$$\approx$0.2--0.4\%.
With the  the standard Coulomb interaction
($e_{\text{MF}}=e$),
the BR isospin mixing increases to  about 0.2--0.7\%.

\begin{figure}
\includegraphics[angle=0,width=0.7\columnwidth]{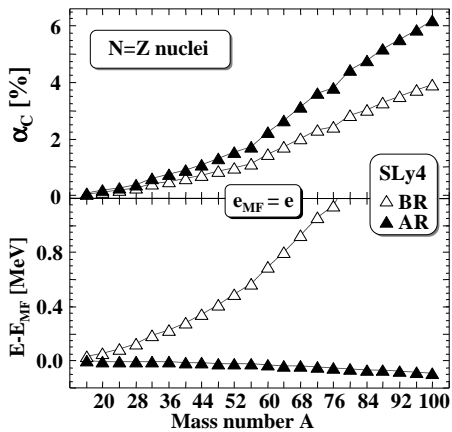}
\caption[T]{\label{fig2}
Upper panel: isospin-mixing parameter $\alpha_C$ in even-even  $N$=$Z$ nuclei
calculated in BR and AR variants. Lower panel:
total $E_{T=0,T_z=0}$ (BR) and $E_{n=1,T_z=0}$ (AR) energies relative to the  MF (or HF) energy $E_{\text{MF}}$.
}
\end{figure}
The AR results are entirely different.
In $^{40}$Ca, with  $e_{\text{MF}}=e$ we
obtain the isospin mixing of 0.9\%, which is about 50\% larger than
the BR value. A similar  increase is predicted for other  $N$=$Z$
systems (see the upper panel of Fig.~\ref{fig2}). This
result nicely illustrates  the
non-perturbative character of the Coulomb polarization when it comes to the isospin mixing.
%
%Indeed, even if without Coulomb interaction there is no spurious isospin mixing in the $N=Z$ systems, it nevertheless severely pollutes the results obtained when the Coulomb interaction is included. This is so, because the spurious-isospin-mixing effect sets in as soon as the Coulomb interaction makes the neutron and proton single-particle states different from one another.
%
The impact of  the isospin mixing on the g.s.\ structure of $N$=$Z$ nuclei
also shows up for the total binding energy (see the lower panel of
Fig.~\ref{fig2}). Differences between the BR and AR energies rapidly
increase with  mass number, to attain about 2\,MeV
in $A$=100. Interestingly, the AR values are amazingly
close to the HF energies $E_{\text{MF}}$, up to
90\,keV. This is  a typical effect of the variational method: the minimum
of energy is reasonably reproduced even if the trial
 wave function is rather incorrect in its detailed structure.

\begin{figure}
\includegraphics[angle=0,width=0.7\columnwidth]{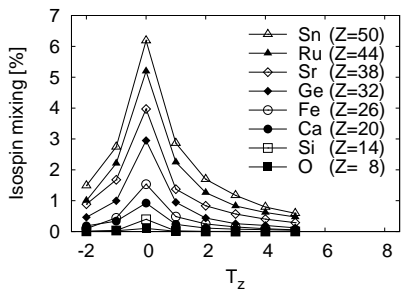}
\caption[T]{\label{fig3}
The AR isospin-mixing parameter $\alpha_C$ calculated
for even-even nuclei with 8$\le$$Z$$\le$50
and $-2$$\le$$T_z$$\le$5.
}
\end{figure}
The AR isospin mixing is rapidly quenched  with
$|N-Z|$. Indeed, $\alpha_C$ in
a $T_z$=1 nucleus
$^{42}$Ca drops to 0.2\%, and then decreases
exponentially to about 0.01\% in $^{60}$Ca. As seen in  Fig.~\ref{fig3}, this
behavior holds for all isotopes. It is
interesting to see in Fig.~\ref{fig1} that the AR results obtained for $e_{\text{MF}}=0$
and $e_{\text{MF}}=e$ are  quite similar beyond $^{44}$Ca. This indicates
that the correct good-isospin basis $|T,T_z\rangle$ is generated
from isospin-broken HF states even without the Coulomb interaction included on a MF level.

\begin{figure}
\includegraphics[angle=0,width=0.7\columnwidth]{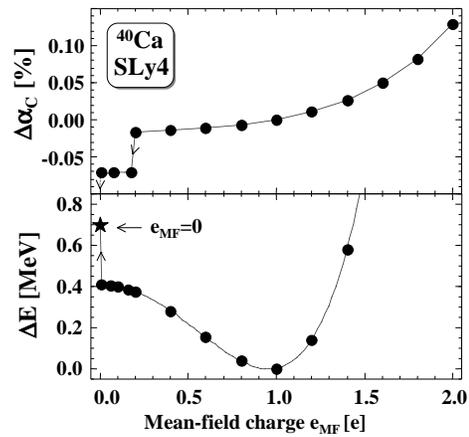}
\caption[T]{\label{fig4}
The isospin-mixing parameter $\Delta\alpha_C$ (upper
panel) and total energy $\Delta E$ (lower panel)  in the AR method  relative to the values obtained with the full Coulomb term, plotted for $^{40}$Ca as a  function of $e_{\text{MF}}$.
}
\end{figure}
This fact is further corroborated by the AR results, shown in
Fig.~\ref{fig4}
relative to those obtained with the full Coulomb term, as a  function of $e_{\text{MF}}$.  For
$0.2\,e\leq{}e_{\text{MF}}\leq{}e$, the isospin mixing of 0.9\%
obtained with $e_{\text{MF}}=e$ does not vary by more than 0.01\%.
At $e_{\text{MF}}=0.2\,e$, the amplitude  of the $|T=2,T_z=0\rangle$ component in the MF wave function becomes too small to be included in the AR
calculation; hence,  the isospin mixing jumps by 0.06\%. This is so,
because at this small value of
$e_{\text{MF}}$,
the $\Delta{T}=2$ coupling of the Coulomb force
becomes ineffective and the numerical accuracy cannot be controlled.
The $|T=1,T_z=0\rangle$ component becomes very small only in the very close
neighborhood of  $e_{\text{MF}}=0$, at which point the
isospin mixing disappears altogether.

The lower panel of Fig.~\ref{fig4} shows the total AR energy as a
function of $e_{\text{MF}}$. Here, we can understand the role of
$e_{\text{MF}}$ as a variational
parameter that can be used to optimize the good-isospin basis
$|T,T_z\rangle$.   It is
gratifying to see that the minimum of energy is obtained almost
exactly at the physical value of $e_{\text{MF}}=e$. Namely, the
optimal wave functions  $|T,T_z\rangle$ are generated
by taking at the MF level the full Coulomb
interaction having the physical charge. However, it is to be noted that
the energy differences in Fig.~\ref{fig4} are
quite small, of the order of a few hundred  keV. Moreover, as
discussed above, the isospin mixing is almost insensitive to such a
refinement of $|T,T_z\rangle$. This result supports our initial assumption: the good-isospin states
$|T,T_z\rangle$ are fairly robust to the variations of the isospin-breaking
interaction, i.e., they well capture
self-consistent polarization effects.

\begin{figure}
\includegraphics[angle=0,width=0.7\columnwidth]{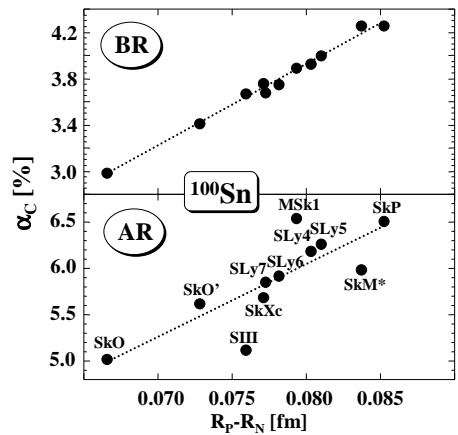}
\caption[T]{\label{fig5}
Isospin-mixing parameter $\alpha_C$ in
$^{100}$Sn BR (upper panel) and
AR (lower panel)
 for various Skyrme EDF parameterizations  as a function of the
difference between the MF proton and neutron rms radii.
Straight lines, representing linear fits,  are drawn  to
guide the eye. See Ref.~\protect\cite{[Ben03s]} for
details of Skyrme functionals used.
 }
\end{figure}

Although our results give first reliable estimates of the isospin
mixing within extended  MF theory, the final values of $\alpha_C$
are still quite uncertain, which is due to an imperfect determination of
the nuclear EDF.
This  is illustrated in Fig.~\ref{fig5}  which shows the  isospin mixing calculated  in BR and AR variants  for a heavy $N$=$Z$ nucleus  $^{100}$Sn
for a wide selection of the Skyrme EDF parameterizations
\cite{[Ben03s]}. We note that  $\alpha_C$  does depend  on the nuclear
effective interaction: the difference between extreme AR values obtained for
SkO and SkP is as large as  1.5\%, which is
about 30\% of the value of the isospin mixing in $^{100}$Sn.

In trying to pin down
those features of the EDF that would be responsible for differences
in $\alpha_C$, we have attempted to find correlations between
isospin mixing  and various EDF characteristics \cite{[Ben03s]}.
We conclude that no clear correlations exist between $\alpha_C$
and those EDF parameters that are related to  nuclear-matter properties.
In particular, this is  true for the  nuclear-matter symmetry energy,
the prime suspect to influence the properties of the isovector channel.
We did find a very clear correlation of the BR values of $\alpha_C$
with the differences between the MF proton and neutron rms radii (see
Fig.~\ref{fig5}).
This is not surprising, as the  monopole polarization does impact the
proton and neutron radii, and their difference.
However, after the rediagonalization, the values of $\alpha_C$ show a much weaker correlation.
Clearly, the precise values of the isospin mixing parameter depend
on fine details of the nuclear EDF.

In conclusion, we performed the self-consistent analysis of isospin
mixing within the extended mean-field approach. Our method is
non-perturbative; it fully takes into account long-range polarization
effects associated with the Coulomb force and neutron excess. The
nuclear Hamiltonian, including the full Coulomb interaction, is
diagonalized in a good-isospin basis obtained by isospin projection from
self-consistent HF states. Not surprisingly, the largest
isospin-breaking effects have been predicted for $N$=$Z$ nuclei, where
the effects due to the neutron (proton) excess are smallest and the
Coulomb force dominates the picture.

The unphysical isospin violation
caused by the neutron excess is significant on the MF level: the largest
effect is predicted in $|N-Z|$=2 nuclei. However, the
rediagonalization procedure eliminates the spurious isospin mixing  almost
completely. While the optimal many-body solutions are obtained by using
the double variational approach, we have demonstrated that one obtains a
reasonable good-isospin basis by broadly varying the strength of the
Coulomb interaction of the EDF. Finally, we investigated the dependence of
isospin mixing on the self-consistent feedback between the nuclear
and Coulomb terms. We found an appreciable dependence of
$\alpha_C$ on  the parametrization of the nuclear functional and found a rough
correlation between the isospin mixing and the difference between proton
and neutron rms radii. The microscopic approach described in this study
will be applied to isospin symmetry-breaking corrections to superallowed
Fermi beta decays, Coulomb energy differences, and  properties of
analogue  states.

Discussions with Erich Ormand are gratefully acknowledged.
This work was supported in part by the Polish Ministry of Science
under Contract No.~N~N202~328234, Academy of Finland and University
of Jyv\"askyl\"a within the FIDIPRO programme, and U.S.~Department of Energy under Contract
Nos.~DE-FG02-96ER40963 (University of Tennessee) and  DE-AC05-00OR22725
with UT-Battelle, LLC (Oak Ridge National Laboratory).

%\bibliography{C:/Actual/LaTeX/Temp/jd,C:/Actual/LaTeX/Latex.all/jacwit25}
%\bibliography{jd,jacwit25}
%\bibliography{jd, C:/pallad/bibfiles/jacwit25}
%\bibliography{jd,C:/wojtek/bibfiles/jacwit25}
%\bibliographystyle{unsrt}

\end{document}